\documentclass[aps,prl,showpacs,amssymb,nofootinbib,twocolumn]{revtex4} 

\usepackage{bm}
\usepackage{amsmath}
\usepackage{amssymb}
\usepackage{latexsym}
\usepackage{amsfonts}
\usepackage{epsfig}
\usepackage{psfrag}
\usepackage{graphicx}

\newcommand{\nn}{\nonumber\\}

\newcommand{\f}[1]{\mbox{\boldmath$#1$}}

\newcommand{\vau}{\mbox{\boldmath$v$}}

\newcommand{\bea}{\begin{eqnarray}}
\newcommand{\ea}{\end{eqnarray}}
\newcommand{\eea}{\end{eqnarray}}
\newcommand{\ord}{{\cal O}}

\begin{document}

\title{On the origin of the particles in black hole evaporation} 

\author{Ralf Sch\"utzhold$^{1,2,*}$ and William G.~Unruh$^{3,+}$}

\affiliation{
$^1$Fachbereich Physik, Universit\"at Duisburg-Essen, 
D-47048 Duisburg, Germany
\\
$^2$Institut f\"ur Theoretische Physik,
Technische Universit\"at Dresden, D-01062 Dresden, Germany
\\
$^3$Canadian Institute for Advanced Research Cosmology and Gravity 
Program
\\
Department of Physics and Astronomy, University of British Columbia, 
Vancouver B.C., V6T 1Z1 Canada
} 

\begin{abstract}
We present an analytic derivation of Hawking radiation for an
arbitrary (spatial) dispersion relation $\omega(k)$ as a model for
ultra-high energy deviations from general covariance. 
It turns out that the Hawking temperature is proportional to the
product of the group $d\omega/dk$ and phase $\omega/k$ velocities
evaluated at the frequency $\omega$ of the outgoing radiation far
away, which suggests that Hawking radiation is basically a low-energy
phenomenon. 
Nevertheless, a group velocity growing too fast at ultra-short
distances would generate Hawking radiation at ultra-high energies
(``ultra-violet catastrophe'') and hence should not be a realistic
model for the microscopic structure of quantum gravity.  
\end{abstract}

 
\maketitle

{\em Introduction}\quad
%
In the history of physics, unexplained coincidences were often the 
precursor of striking discoveries.
For example, the limiting propagation velocity derived from the
properties of coils and capacitors in table-top experiments turned out
to be strikingly close to the speed of light measured via planetary
motion -- which guided the unification of these phenomena by the laws
of electrodynamics and ultimately lead to the theory of relativity. 
Today, a similar riddle is the question of {\em why} black holes seem 
to behave like thermal objects \cite{thermo} and evaporate by emitting
Hawking radiation \cite{hawking,unruh,point}.  

One way of achieving a better understanding of these links is to study
the origin of the particles in black hole evaporation, i.e., the
question of {\em where} they are created.
For example, it has been suggested to resolve the black hole 
``information paradox'' (i.e., the apparent contradiction between
unitarity and the second law of thermodynamics in this system) 
by encoding information into the outgoing Hawking particles.
Clearly, this (hidden) encoding mechanism should then occur at
the origin of the particles (or on their way to infinity). 

Since the event horizon marks the ``point of no return'', the modes 
containing the Hawking particles originate from the region very close
to the horizon, i.e., from very short wavelengths 
(gravitational red-shift). 
However, the origin of the {\em modes} is not necessarily the place
where the {\em particles} are created.

In order to address this problem, we derive Hawking radiation in the 
presence of a very general dispersion relation 
$\omega=ck\to\omega=\omega(k)$ associated to the propagating degrees
of freedom.
Such a modified dispersion relation is inspired by the analogy to 
condensed matter, i.e., the sonic black hole analogues 
(silent or ``dumb'' holes), which rely on the quantitative analogy
between quantum fields in curved space-times and phonons 
(or other quasi-particles) propagating in fluids with a general flow
velocity $\vau(t,\f{r})$, cf.~\cite{unruh-prl,jacobson}. 
%
In this case, the phase $\omega/k$ and group  $d\omega/dk$ velocities
vary with wavelength and thus the dependence of Hawking radiation on
the dispersion relation should show us which wavenumbers $k$ are most
important for particle creation: 
The emitted radiation at various frequencies $\omega$ will ``see'' 
different horizons. 
The question this paper will try to answer is what determines the
temperature of the radiation emitted at any particular frequency
$\omega$. 
If the temperature is determined when the wavelengths are very small, 
and frequencies large -- i.e., when the horizon first splits the
incoming wave packet (in its vacuum state) into positive pseudo-norm
modes outside the horizon (which will turn into the Hawking particles) 
and negative pseudo-norm modes (their infalling partner particles)
inside -- then one would expect a universal temperature for all of the
low-frequency modes. 
On the other hand, if it is the low frequency aspects of the modes
which determine the temperature, one might expect the properties of
the horizon defined via one of the velocities (phase, group, or other)
associated with the wave at low frequencies to dictate the temperature
(via $dv/dr$ at that horizon location) of that mode.


Previous analytic calculations were restricted to low energies
$\omega$ and a small vicinity of horizon, see, e.g.,
\cite{universality}, while numerical studies necessarily involved 
a given (mostly sub-luminal) dispersion relation within a restricted
parameter range. 
In the following, we present an analytic derivation of Hawking
radiation valid for any frequency $\omega$ and almost arbitrary
(spatial) dispersion relations $\omega(k)$; the only assumption we
make is that the black or ``dumb'' hole should be large, i.e.,
macroscopic, and that the velocity profile of the background flow have 
a specific form.  

{\em Dispersion relation}\quad
%
Since Hawking/``dumb''-hole radiation is basically a 1+1 dimensional
effect, we consider the Painlev{\'e}-Gullstrand-Lema{\^\i}tre metric 
in 1+1 dimensions ($\hbar=c=G_{\rm N}=k_{\rm B}=1$)
\bea
ds^2=\left[1-v^2(x)\right]dt^2-2v(x)\,dt\,dx-dx^2
\,,
\ea
%
with the local frame dragging velocity $v(x)$, which
corresponds to the flow velocity of the fluid analogue
\cite{unruh-prl}.  
It determines the position of the horizon via $v(x)=\pm c$ where $c$
is the velocity of the propagating modes, which is assumed to be
constant and set to unity here. 
The propagation of a massless scalar field $\Phi$ in this metric 
(or, equivalently, phonons in the fluid) is governed by the
d'Alembertian  
\bea
\label{box}
\Box\Phi
&=&
\left((\partial_t+\partial_xv)(\partial_t+v\partial_x)-\partial_x^2\right)
\Phi
\nn
&=&
(\partial_t+\partial_xv+\partial_x)(\partial_t+v\partial_x-\partial_x)\Phi 
\,.
\ea
Due to the conformal invariance in 1+1 dimensions, the left-moving modes 
$(\partial_t+\partial_xv-\partial_x)\Phi=0$ propagating against the 
frame-dragging velocity are decoupled from the right-moving solutions 
$\phi=(\partial_t+\partial_xv-\partial_x)\Phi$.
However, an arbitrary modification of the dispersion relation would
not preserve this decoupling in general. 
The choice 
\bea
\label{arbitrary}
\Box_h
=
(\partial_t+\partial_xv)(\partial_t+v\partial_x)-
h(\partial_x^2) 
\,,
\ea
for example, does not factorize for a general function $h$. 
On the other hand, if we modify the left and right-moving branch 
separately with an arbitrary function $f$ via 
\bea
\label{separately}
\Box_f
&=&
\left(\partial_t+\partial_x[1+v+f(-\partial_x^2)]\right)
\times
\nn
&&
\left(\partial_t-[1-v+f(-\partial_x^2)]\partial_x\right)
\,,
\ea
the left-moving modes 
$(\partial_t-[1-v+f(-\partial_x^2)]\partial_x)\Phi=0$
are again decoupled from the right-moving solutions given by 
$\phi=(\partial_t-[1-v+f(-\partial_x^2)]\partial_x)\Phi$.  
The difference between the two options (\ref{arbitrary}) and 
(\ref{separately}) scales with the derivative of $v(x)$ compared with
the characteristic scale of the dispersion relation 
$[v(x),f(-\partial_x^2)]$, which is negligibly small in the
hydrodynamic limit (which corresponds to the black hole being large). 
The coupling occurs 
at high frequencies over long scales (the scale of variation of $v$)
which means the response will be adiabatic and left-moving waves will
not convert to right-moving.  
This is the reason why the numerical simulations using the first
option (\ref{arbitrary}) never saw a mixing between left and
right-movers one might expect from the coupling between the two sets
of modes.  

\begin{figure}[ht]
\includegraphics[height=3cm]{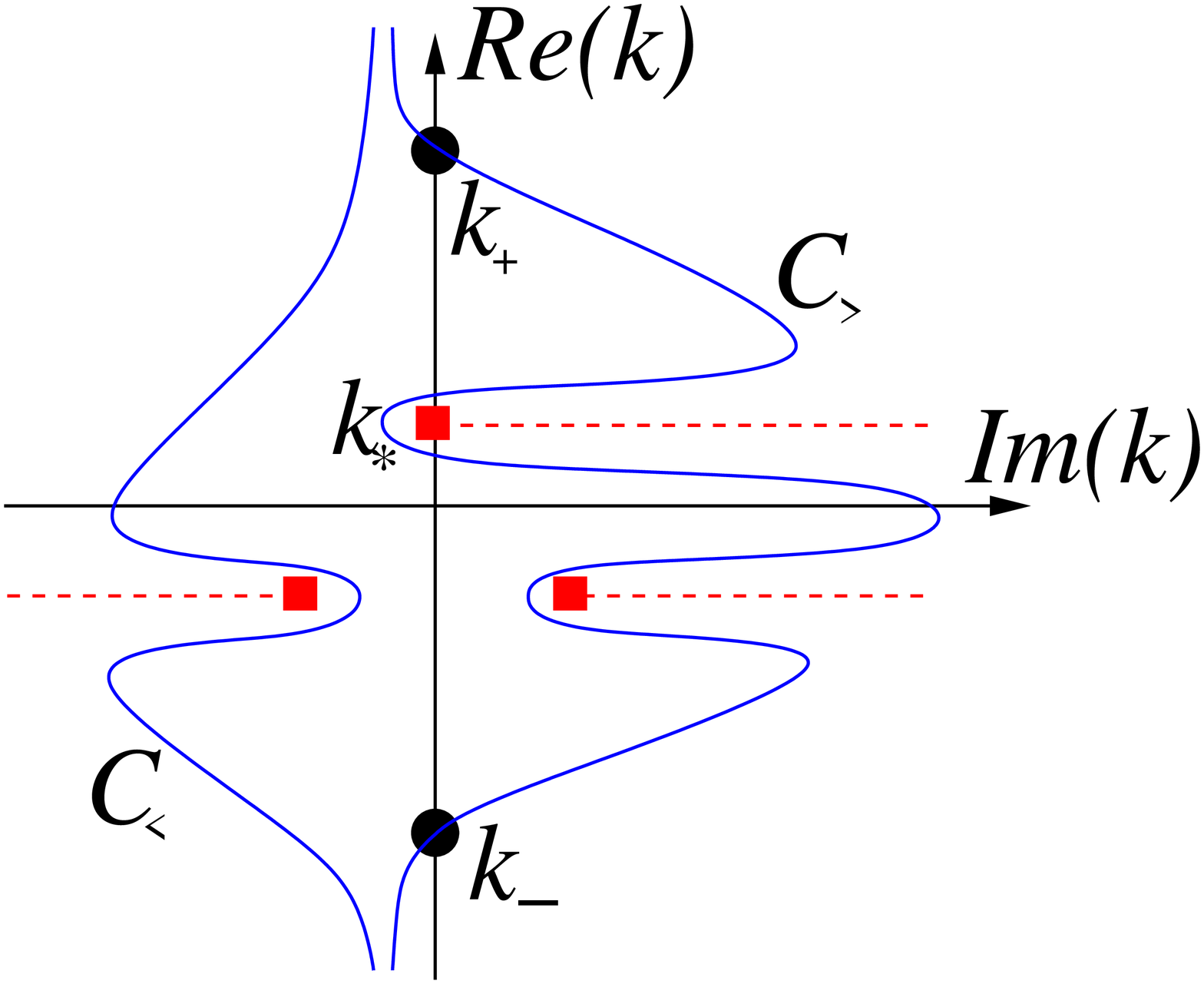}
\hspace{.3cm}
\includegraphics[height=3cm]{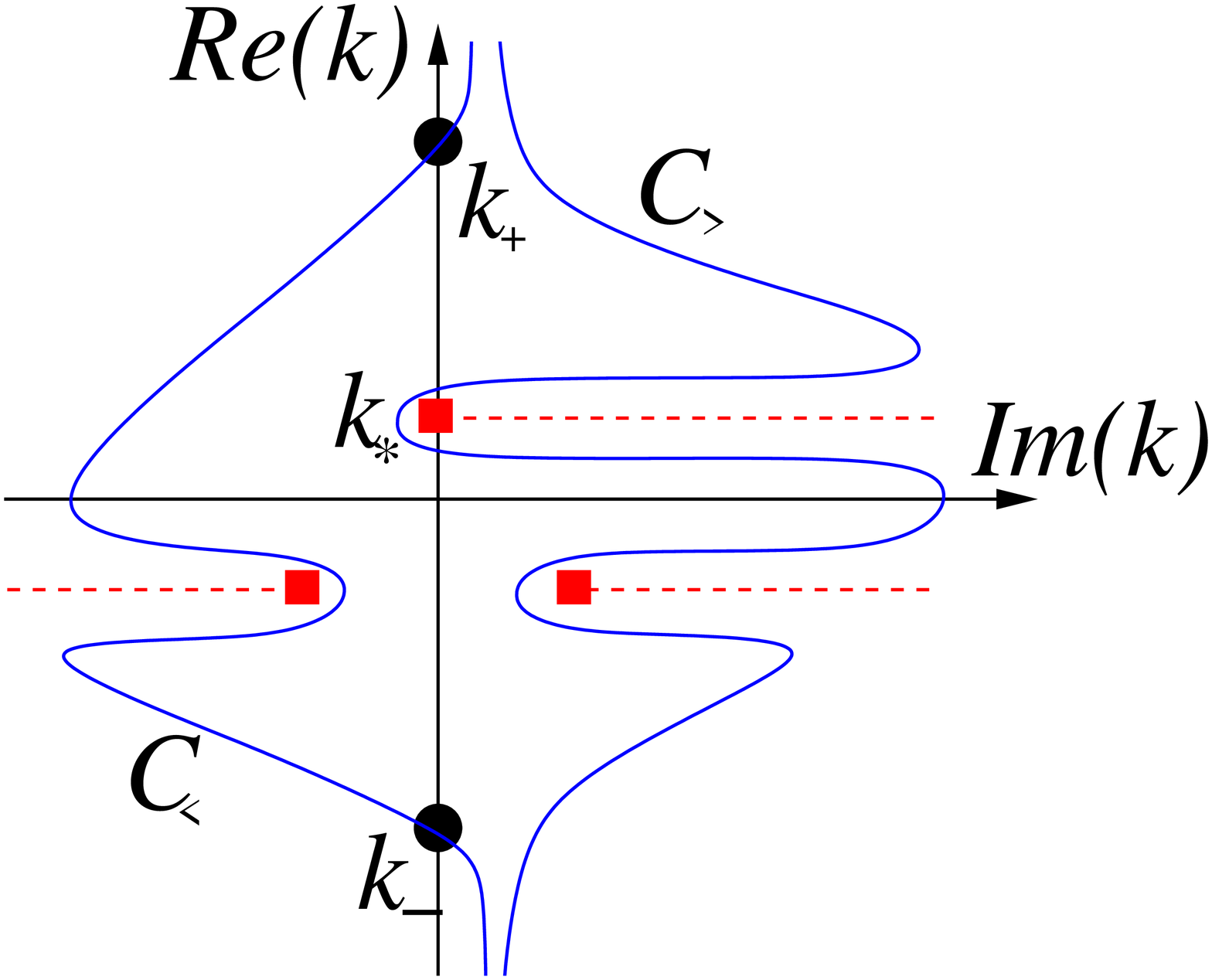}
\caption{\label{figure} [Color online] Sketch (not to scale) of
  integration contours in the complex plane for the sub-luminal (left)
  and the super-luminal case (right). The dots denote saddle points
  $k_\pm$ at the real axis and the squares are the singularities
  $k_\alpha$ with the associated branch cuts (dashed lines).}
\end{figure}

{\em Derivation}\quad
%
Even though the Schwarzschild metric corresponds to
$v(x)=\pm\sqrt{2M/x}$, we consider two different velocity profiles 
$v(x)=-\lambda/x$ and $v(x)=\kappa x$ in the following, because they
admit analytic solutions. 
The latter of course has no asymptotically flat region, and calling
it a black, or ``dumb'', hole is metaphorical. 
Let us first study the case $v(x)=-\lambda/x$. 
After a Fourier-Laplace transformation with $\partial_x\to ik$ and 
$x\to i\partial_k$ as well as $\partial_t\to-i\omega$, 
the left-moving solutions satisfy the integral equation 
(since $\partial_k^{-1}=\int dk$)
\bea
\left(\omega-k\left[1+i\lambda\partial_k^{-1}+f(k^2)\right]\right)
\phi_\omega(k)=0 
\,,
\ea
which can be solved via separation of variables and differentiation 
$\phi_\omega(k)=\exp\{-i\lambda\int dk'/g(k')\}/g(k)$
%
%
with the spectral function $g(k)=1+f(k^2)-\omega/k$. 
%
%
The inverse Fourier-Laplace transformation 
\bea
\label{inverse}
\phi_\omega(r)=\int\frac{dk}{g(k)}\,
\exp\left\{ikx-i\lambda\int\frac{dk'}{g(k')}\right\}
\,,
\ea
yields the spatial modes $\phi_\omega(x)$ with the integration contour
being determined by the boundary conditions. 
In the following, we shall assume the length scale $\lambda$ on
which $v(x)$ changes to be very large compared with the typical
wavenumber of the dispersion relation $f(k^2)$ and the frequency
$\omega$. 
In terms of the fluid analogue, this is precisely the hydrodynamic 
limit -- whereas, for real black holes, is corresponds to demanding
that the size of the black hole is much larger than the Planck scale.  
Note, however, that we do not restrict $\omega$ relative to the Planck
scale (i.e., $\omega$ could be Planckian).  
Since the exponent in the integral above contains the large numbers
$x$ and $\lambda$, it is very useful to deform the integration contour
into the complex plane (assuming that $f$ is an analytic function),  
where the leading contributions will be determined by singularities
and the associated branch cuts as well as saddle points.
The saddle points (stationary phase) 
\bea
\label{stationary}
x=\frac{\lambda}{g(k)}
\,\leadsto\,
\omega=k\left[1-\frac{\lambda}{x}+f(k^2)\right]
\,,
\ea
are solutions of dispersion relation  
$\omega+vk=k[1+f(k^2)]$. 
For large wavenumbers $|k|\gg\omega$, we get pairs of saddle points
$k_\pm$ satisfying $f(k^2_\pm)=\lambda/x-1$.  
The singularities of the integrand at $g(k)=0$ correspond to solutions
of dispersion relation far away from the black hole $x\to\infty$.
Assuming simple poles at $k_\alpha$ only, we may employ the residual 
expansion  
\bea
\label{residual}
\frac{1}{g(k)}=\sum\limits_\alpha\frac{c_\alpha}{k-k_\alpha}
\,\leftrightarrow\,
c_\alpha=\frac{1}{2\pi i}\oint\limits_{{\mathfrak C}_\alpha} 
\frac{dk}{g(k)}
=\frac{k_\alpha}{v_{\rm gr}(k_\alpha)}
\,.
\ea
The contours ${\mathfrak C}_\alpha$ denote small circles around the
poles at $k_\alpha$ and the residual coefficients $c_\alpha$ are 
related to the group velocity at these points. 
Insertion of the residual expansion (\ref{residual}) into
Eq.~(\ref{inverse}) yields 
\bea
\phi_\omega(x)=\int\frac{dk}{g(k)}\,e^{ikx}\,
\prod\limits_\alpha\left(k-k_\alpha\right)^{-i\lambda c_\alpha}
\,.
\ea
Consequently, there are branch cuts starting from each singularity
unless $2iMc_\alpha\in\mathbb Z$, cf.~Fig.~\ref{figure}.  

{\em Sub-luminal case}\quad
%
In order to determine the most suitable integration contour in the
complex plane, we have to incorporate some assumptions about the
function $f(k^2)$. 
First, we assume that the asymptotic ($x\to\infty$) dispersion relation
$\omega=k[1+f(k^2)]$ is convex, i.e., sub-luminal $f(k^2)<0$, and 
always monotonically increasing, i.e., with a positive group velocity.  
In this case, saddle points in Eq.~(\ref{stationary}) occur outside
the horizon $x>\lambda$ and 
yield the following contributions to the integral in
Eq.~(\ref{inverse}) 
\bea
\label{saddle}
\phi_\omega^\pm(x)\approx\sqrt{\frac{2i\pi}{\lambda g'(k_\pm)}}\,
e^{ik_\pm x}\, 
\prod\limits_\alpha\left(k_\pm-k_\alpha\right)^{-i\lambda c_\alpha}
\,. 
\ea
In view of their spatial behavior, these are the positive
$\phi_\omega^+$ and negative $\phi_\omega^-$ pseudo-norm in-modes with
large wavenumbers $|k_\pm|\gg\omega$ and therefore small group
velocities, which are swept towards the horizon.  
After quantizing the field $\Phi$, these positive/negative pseudo-norm  
solutions yield the creation/annihilation operators of the in-modes at
large $k$.

In order to close the integration contour, we have to circumvent the
branch cuts in the upper complex half plane $\Im(k)>0$,
cf.~Fig.~\ref{figure}.   
In the limit $x\to\infty$, the contributions of the branch cuts
starting from the singularities $k_\alpha$ away from the real axis
$\Im(k_\alpha)>0$ are exponentially suppressed and only the branch cut
starting at the real axis $\Im(k_*)=0$ contributes. 
This wavenumber $k_*$ represents a real solution of the dispersion
relation far away from the horizon and just corresponds to the
outgoing Hawking radiation with frequency $\omega>0$. 
Hence, at $x\to\infty$, the integral in Eq.~(\ref{inverse})
corresponding to the contour ${\mathfrak C}_>$ in Fig.~\ref{figure} 
yields a superposition of the outgoing Hawking modes with $k_*$ and
the large-$k$ in-modes in (\ref{saddle}) with $k_\pm$. 
Continuing this solution beyond the horizon $x<\lambda$, the saddle
points vanish and the integrand in (\ref{saddle}) decays exponentially
in the lower complex half plane $\Im(k)<0$. 
Thus, we deform the integration contour to ${\mathfrak C}_<$, where
the main contribution stems from the branch cut(s) starting at 
$\Im(k_\alpha)<0$. 
Again, for large $\lambda$, these contributions are exponentially
suppressed.  

Therefore, the outgoing Hawking mode with $k_*$ originates entirely
from the in-modes in (\ref{saddle}).  
Assuming that the initial quantum state is the ground state of the 
large-$k$ modes in (\ref{saddle}) with respect to the freely falling
observer, the amount of created particles is then determined by the
mixture between these positive and negative pseudo-norm solutions with
large $k$ contained in the outgoing $k_*$-modes after the immense
gravitational red-shift near the horizon.  
In view of $|k_\alpha|\ll|k_\pm|$, the only difference between
positive and negative pseudo-norm modes is caused by branch 
cut(s) 
\bea
\label{ratio}
\left|\frac{\phi_\omega^+(x)}{\phi_\omega^-(x)}\right|\approx
\exp\left\{\pi\lambda\sum\limits_\alpha (-1)^{s_\alpha} 
\Re(c_\alpha)\right\}
\,,
\ea
where $(-1)^{s_\alpha}$ is the sign associated with the direction of
the branch cut.
Since $g(k)$ is a real function, the singularities $k_\alpha$ occur 
symmetric w.r.t.\ the real axis $k_\alpha\to k_\alpha^*$ and 
$c_\alpha\to c_\alpha^*$. 
Choosing the branch cuts suitably (see Fig.~\ref{figure}), the 
contributions from the symmetric pairs cancel each other and 
hence only the singularity at the real axis 
$k_*=k_\alpha\in\mathbb R$ contributes. 
Together with the unitarity relation 
$|\alpha_\omega|^2-|\beta_\omega|^2=1$,
the ratio (\ref{ratio}) directly determines the size of the
Bogoliubov coefficients via 
$|\alpha_\omega/\beta_\omega|^2=|\phi_\omega^+/\phi_\omega^-|^2
=\exp\{\omega/T\}$. 
Hence we may read off the effective Hawking temperature  
\bea
\label{temperature}
T_{\rm Hawking}(\omega)
=
\frac{v_{\rm gr}(k_*)v_{\rm ph}(k_*)}{2\pi\lambda}
\,.
\ea
We observe that the geometric mean of group and phase velocity
\cite{mean} evaluated at the frequency $\omega$ of the outgoing
radiation far away $x\to\infty$ determines the Hawking temperature
\cite{triest}. 
Therefore, the behavior of the dispersion relation at large $k$ is
not relevant -- even though the Hawking radiation originates from
large-$k$ modes -- which indicates that the Hawking effect is
basically a low-energy phenomenon. 
The $\omega$-dependence of the Hawking temperature can be explained by
the fact that high-energy wave-packets have a different group velocity
than those at low energy and hence the various modes ``see'' different 
horizons and thus other values for the surface gravity, i.e., 
velocity gradient $dv/dx=\lambda/x^2\propto v^2/\lambda$. 

This explanation has been confirmed by numerical simulations 
\cite{triest} and can further be supported by considering the second
case $v(x)=\kappa x$, where the velocity gradient $\kappa$ is
constant. 
In this case, Eq.~(\ref{inverse}) should be replaced by 
\bea
\phi_\omega(x)=\int dk\,k^{-i\omega/\kappa}
\exp\left\{ikx-i\kappa\int dk'f(k'^2)\right\}
\,.
\ea
Hence the weight of the branch cut starting at $k=0$ is just
determined by the ratio $\omega/\kappa$ and the Hawking temperature
does not depend on $\omega$ at all because all modes ``see'' the same
surface gravity $\kappa$ 
\bea
\label{kappa}
T_{\rm Hawking}
=
\frac{\kappa}{2\pi}
=
\rm const
\,.
\ea
%

{\em Super-luminal case}\quad
%
A super-luminal dispersion relation $f(k^2)$ can be treated in a
completely analogous way.
As the only difference, the large-$k$ in-modes determined by the
saddle points $k_\pm$ originate from inside the horizon $x<\lambda$
and thus the contours in the complex plane 
(${\mathfrak C}_>$ for $x>\lambda$ and ${\mathfrak C}_<$ for
$x<\lambda$, cf.~Fig.~\ref{figure}) are slightly different.

In view of this observation, one might wonder whether the ansatz 
$v(x)=-\lambda/x$ instead of $v(x)=\pm\sqrt{2M/x}$ is justified. 
To address this question, let us consider the Schwarzschild geometry
in 1+1 dimensions using the Eddington-Finkelstein coordinates $(V,r)$ 
\bea
ds^2=\left(1-\frac{2M}{r}\right)dV^2-2dV\,dr
\,.
\ea
For a massless scalar field $\Phi$, the wave equation reads 
\bea
\left(2\partial_V\partial_r+
\partial_r\left[1-\frac{2M}{r}+f(-\partial_r^2)\right]\partial_r
\right)\Phi=0
\,,
\ea
where we have again included a modification $f(k^2)$ of the dispersion 
relation. 
Comparison with the previous derivation yields completely the same
results for the outgoing solutions $\phi=\partial_r\Phi$ up to the
replacement $\omega\to2\omega$ due to the Eddington-Finkelstein
coordinate $t \to V$. 

However, in the super-luminal case, an additional complication may
arise: 
According to Eqs.~(\ref{residual}) and (\ref{ratio}), the thermal
Boltzmann factor $\exp\{\omega/(4T)\}$ determining the amount of
created particles with frequency $\omega$ can be recast into 
the alternative form $\exp\{2\pi M k_*/v_{\rm gr}(k_*)\}$.  
Hence, if the group velocity grows slower than linear in $k$,  
the number of produced particles decreases with energy.  
However, if $v_{\rm gr}(k)$ rises too fast in some $k$-region, the
amount of created particles drops at low $k$ (where $v_{\rm gr}=1$)
but later increases again!
In such an extremal case, the Hawking radiation could contain a large  
contribution of ultra-high energy particles 
(``ultra-violet catastrophe'').  
Going a step further and taking the dispersion relation seriously as a
model for ultra-high energy deviations from general relativity
\cite{metric}, one would exclude such a case in view of our
observational evidence for the existence of black holes with
macroscopic life-times.  

Let us discuss some examples:
The dispersion relation $\omega^2=k^2+k^4$ which is realized for the
sonic black-hole analogues in Bose-Einstein condensates \cite{garay},
does not generate an ``ultra-violet catastrophe'' and reproduces
Hawking's prediction. 
In contrast, the expressions $\omega^2=\exp\{k^2\}-1$ or 
$\omega=k/\sqrt{1-k^2}$ grow too fast and hence lead to the
aforementioned problems \cite{motiv}.  

Note that the condition $v_{\rm gr}(k_*)\geq\ord(Mk_*)$ for particle
creation obtained from the Boltzmann factor precisely marks the
break-down of the saddle-point (i.e., geometric optics)
approximation. 
Writing the integrand in Eq.~(\ref{inverse}) as $G(k)\exp\{F(k)\}$,
the first-order corrections to the saddle-point expansion scale as 
$G'/G \times F^{(3)}/(F'')^2$, $F^{(4)}/(F'')^2$, and $G''/(GF'')$,
evaluated at the saddle point $F'=0$. 
In our case (\ref{inverse}), we have $F''=-2iMG'$ and hence inserting 
$v_{\rm gr}(k_*)\geq\ord(Mk_*)$ yields ``corrections'' of order one --
i.e., the  saddle-point approximation fails. 
Even if the modes started out in their ground state (at $k_\pm$), they 
get excited (at $k_*$) due to the strong gravitational red-shift. 
Based on these general adiabaticity arguments, one would expect that
the main result remains valid even beyond a sole modification of the
dispersion relation:
If the outgoing Hawking modes originate from the vicinity of the
singularity and the spectral properties of quantum gravity change too
fast with energy, one would expect a break-down of adiabaticity at
short distances resulting in the emission of high-energy particles. 
This mechanism is not necessarily restricted to a Planck-length
vicinity of the singularity since the effective surface gravity
``seen'' by the high-energy modes scales with $2M/r^2$ and hence may
exceed the Planck temperature already many Planck lengths away from
the singularity. 

{\em Acknowledgments}\quad
%
This work was supported by the Emmy-Noether Programme of the German
Research Foundation (DFG) under grant \# SCHU~1557/1-2,3; as well as 
the Canadian Institute for Advanced Research; and the
Natural Science and Engineering Research Council of Canada. 
%
\\
$^*${\tt schuetz@theory.phy.tu-dresden.de}
\\
$^+${\tt unruh@physics.ubc.ca}



\begin{thebibliography}{499}
\bibitem{point}
A potentially related point is the connection between acceleration and  
temperature due to quantum effects \cite{unruh}.
%
Of course, we can do the math and derive this connection -- but this
does not mean that we fully understand the underlying reason for it.
\bibitem{mean}
Note that the geometric mean of group and phase velocity for a massive
particle with $\omega^2=c^2k^2+m^2c^4$ yields $c$, while the
arithmetic or harmonic mean, for example, diverges at $k\to0$.
\bibitem{metric}
Of course, one should also ask the question of whether the assumed
$1-2M/r$ behavior of the metric is still valid is this extremal
regime.
\bibitem{motiv}
In order to motivate the third example, one could argue that it should 
be impossible to resolve spatial details smaller than the Planck
length 
$\ell_{\rm P}=\sqrt{\hbar G_{\rm N}/c^3}\approx1.6\times10^{-35}$m,
because Heisenberg's uncertainty principle would then require
concentrating a large (trans-Plankian) energy and thus mass in a small 
volume -- which should form a black hole. 
%
In analogy to special relativity, where the acceleration of massive
particles towards the speed of light is prohibited by a diverging
energy $E=mc^2/\sqrt{1-v^2/c^2}$, one could speculate that quantum
gravity impedes a spatial resolution better than the Planck length
$\ell_{\rm P}$ in a similar manner 
$\omega=ck/\sqrt{1-\ell_{\rm P}^2k^2}$.
%
%
Even though the two relations $\omega^2=k^2+k^4$ and 
$\omega=k/\sqrt{1-k^2}$ can probably not be
distinguished via observations in the low-energy sector 
(e.g., energy-dependent time-delay measurements of signals from
distant galaxies), they induce drastic differences for black holes.
\end{thebibliography}
\end{document}